\documentclass[aps,
    reprint,
    groupedaddress,
    amsmath,
    amssymb
]{revtex4-1}

\usepackage{graphicx}
\usepackage[mathscr]{euscript}
\usepackage{amsmath}
\usepackage{color}
\usepackage{subfig}
\usepackage{array}
\usepackage[lined,boxruled]{algorithm2e}

\begin{document}

\title{A non-parametric $k$-nearest neighbour entropy estimator}

\author{Damiano Lombardi and Sanjay Pant}
\email[]{\{Damiano.Lombardi, Sanjay.Pant\}@inria.fr}
\affiliation{
    Inria Paris-Rocquencourt, BP 105, 78153 Le Chesnay Cedex, France \\
    Sorbonne Universit\'es, UPMC Univ Paris 06, 4 Place Jussieu, 75252 Paris cedex 05, France \\
    CNRS, UMR 7598 Laboratoire Jacques-Louis Lions, Paris, France
}


\date{\today}

\begin{abstract}
A non-parametric $k$-nearest neighbour based entropy estimator is proposed. It improves on the classical Kozachenko-Leonenko estimator by considering non-uniform probability densities in the region of $k$-nearest neighbours around each sample point. It aims at improving the classical estimators in three situations: first, when the dimensionality of the random variable is large; second, when near-functional relationships leading to high correlation between components of the random variable are present; and third, when the marginal variances of random variable components vary significantly with respect to each other. Heuristics on the error of the proposed and classical estimators are presented. Finally, the proposed estimator is tested for a variety of distributions in successively increasing dimensions and in the presence of a near-functional relationship. Its performance is compared with a classical estimator and shown to be a significant improvement.
\end{abstract}


\maketitle

\section{Introduction}

\emph{Entropy} is a fundamental quantity in information theory that finds applications in various areas such as coding theory and data compression \cite{cover2012elements}. It is also a building block for other important measures, such as \emph{mutual information} and \emph{interaction information}, that are widely employed in the areas of computer science, machine learning, and data analysis. In most realistic applications, the underlying true probability density function (pdf) is rarely known, but samples from it can be obtained via data-acquisition, experiments, or numerical simulations. An interesting problem then, is to estimate the entropy of the underlying distribution only from a finite number of samples. The approaches to perform such a task can broadly be classified into two categories: \emph{parametric} and \emph{non-parametric}. In the parametric approach the form of the pdf is assumed to be known and its parameters are identified from the samples. This, however, is a strong assumption and in most realistic cases an \emph{a priori} assumption on the form of the pdf is not justified. Consequently, non-parametric approaches where no such assumption is made have been proposed \cite{beirlant1997nonparametric}. One such approach is to first estimate the pdf through histograms or kernel density estimators (KDE) \cite{silverman1986density,devroye1985nonparametric,scott2015multivariate}, and then to compute the entropy by either numerical or Monte-Carlo (MC) integration. Other alternatives include methods based on sample spacings for one-dimensional distributions \cite{hall1984limit,dudewicz1987empiric} and $k$-nearest neighbours (kNN) \cite{Kozachenko1987sample,tsybakov1996root,singh2003nearest,kraskov2004estimating}.

While KDE based entropy estimation is generally accurate and efficient in low dimensions, the method suffers from the \emph{curse of dimensionality} \cite{gray2003nonparametric}. On the other hand the kNN based estimators are computationally efficient in high dimensions, but not necessarily accurate, especially in the presence of large correlations or functional dependencies \cite{gao2014Efficient}. The latter problem has recently been addressed by estimating the local non-uniformity through principal component analysis (PCA) in \cite{gao2014Efficient}. In the current work, a different approach to overcome the aforementioned limitations associated with kNN based entropy estimators is presented. The central idea is to estimate the probability mass around each sample point by a local Gaussian approximation. The local approximation is obtained by looking at $p$-neighbours around the sample point. This procedure has two distinct advantages: first, that the tails of the true probability distribution are better captured; and second, that if the probability mass in one or more directions is small due to large correlations (near-functional dependencies), or due to significant variation in the marginal variances of the random variable components, the non-uniformity is inherently taken into account. These two features allow the entropy to be estimated in high dimensions with a significantly lower error when compared to classical estimators.

The structure of the work is as follows: first, the classical and the new kNN estimators are presented in section \ref{sec:formulation}; then, the heuristics on the errors of the two estimators are presented in section \ref{sec:heuristics}; and finally, numerical test cases are presented in section \ref{sec:numerical_testcases} for a variety of distributions in successively increasing dimensions.

\section{Formulation of the entropy estimator}
\label{sec:formulation}
%
%
Let the random variable under consideration be $\mathbf{X} \in \mathbb{R}^{d}$ and its probability density be denoted by $p_{\mathbf{X}}(\mathbf{x})$. Its entropy is defined as
\begin{equation}
    H(\mathbf{X}) = \int_\mathcal{X} p_{\mathbf{X}}(\mathbf{x}) \log \left( \frac {1}{p_{\mathbf{X}}(\mathbf{x})} \right) \ d\mathbf{x}
\end{equation}
\noindent
where $\mathcal{X}$ is the support of $p_{\mathbf{X}}(\mathbf{x})$. The goal is to estimate $H(\mathbf{X})$
from $N$ finite samples, $\mathbf{x}_i \ \ i=1 \ldots N$, from the distribution $p_{\mathbf{X}}(\mathbf{x})$ . A Monte-Carlo estimate of the entropy can be written as
\begin{equation}
    \label{eqn:MCestimate}
    \hat{H}(\mathbf{X}) = \frac{1}{N} \sum_{i=1}^{N} \ \log \left( \frac{1}{p_{\mathbf{X}}(\mathbf{x}_i)} \right).
\end{equation}
However, since $p_{\mathbf{X}}(\mathbf{x}_i)$ is unknown, an estimate $\hat{p}_{\mathbf{X}}(\mathbf{x}_i)$ must be substituted in equation \eqref{eqn:MCestimate} to obtain $\hat{H}(\mathbf{X})$.

\begin{figure}[htpb]
    \centering
\includegraphics[width=0.7\columnwidth]{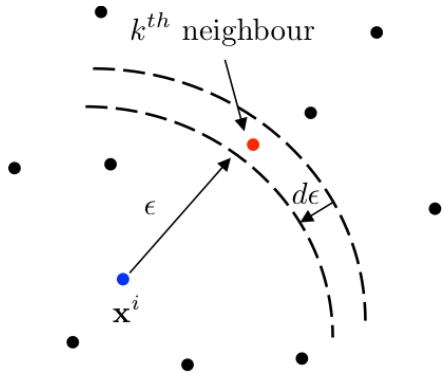}
    \caption{A depiction of $k$-nearest neighbour and $\varepsilon$-ball.}
    \label{fig:kNN_drawing}
\end{figure}

The key idea is to estimate $\hat{p}_\mathbf{X}(\mathbf{x}_i)$ through $k$-nearest neighbours (kNN) of $\mathbf{x}_i$.
%
Consider the probability density $p_k(\varepsilon)$ of $\varepsilon$, the distance from $\mathbf{x}_i$ to its kNN (see Figure \ref{fig:kNN_drawing}).
The probability $p_k (\varepsilon) d\varepsilon$ is the probability that exactly one point is in $[\varepsilon,\varepsilon+d\varepsilon]$, exactly $k-1$ points are at distances less than the kNN, and the remaining points are farther than the kNN. Then it follows that
\begin{equation}
    \label{eqn:combination}
\footnotesize    p_k (\varepsilon)d\varepsilon  = {N-1\choose 1} \frac{dP_i(\varepsilon)}{d\varepsilon} d\varepsilon \ {N-2\choose k-1}
\left( P_i(\varepsilon) \right)^{k-1} \
\left(1-P_i(\varepsilon) \right)^{N-k-1}
\end{equation}
\noindent
where, $P_i(\mathbf{\varepsilon})$ is the probability mass of an $\varepsilon$-ball centered at a sample point $\mathbf{x}_i$. The region inside the $\varepsilon$-ball is $||\mathbf{x}-\mathbf{x}_{i}||<\varepsilon$ and is denoted by denoted by $\mathscr{B}(\varepsilon,\mathbf{x}_{i})$. The probability mass in $\mathscr{B}(\varepsilon,\mathbf{x}_{i})$ is
\begin{equation}
    \label{eqn:mass_eball}
    P_i(\varepsilon)  = \int\displaylimits_{\mathscr{B}(\varepsilon,\mathbf{x}_{i})} p_{\mathbf{X}}(\mathbf{x}) \ d\mathbf{x}.
\end{equation}


%
\noindent
The expected value of $\log(P_i)$ can be obtained from equations \eqref{eqn:combination} and \eqref{eqn:mass_eball}
\begin{equation}
    \label{eqn:expected_probability_mass}
    \mathbb{E} (\log P_i) = \int_0^\infty \log P_i(\varepsilon) \ p_k({\varepsilon}) \ d\varepsilon = \psi(k) - \psi(N)
\end{equation}
\noindent
where $\psi$ is the digamma function.

If the probability mass in $\mathscr{B}(\varepsilon,\mathbf{x}_{i})$ can be written in the following form
\begin{equation}
    \label{eqn:probability_mass_general_form}
    P_i \approx \eta_i \ p_\mathbf{X}(\mathbf{x}_i)
\end{equation}
\noindent
then, by considering the logarithm and taking expectations on both sides of equation \eqref{eqn:probability_mass_general_form}, and using equations \eqref{eqn:expected_probability_mass} and \eqref{eqn:MCestimate}, the entropy estimate can be written as
\begin{equation}
    \label{eqn:generic_entropy_estimate}
    \hat{H}(\mathbf{X}) = \psi(N) - \psi(k) + \frac{1}{N} \sum \log \eta_i.
\end{equation}
In what follows the classical manner to obtain equation \eqref{eqn:probability_mass_general_form} and the new estimator are presented.

\subsection{Classical estimators}
\noindent
The classical estimates by Kozachenko and Leonenko \cite{Kozachenko1987sample,kraskov2004estimating}, and similarly by Singh et. al. \cite{singh2003nearest}, assume that the probability density $p_\mathbf{X}(\mathbf{x})$ is constant inside $\mathscr{B}(\varepsilon,\mathbf{x}_{i})$. For example Kozachenko and Leonenko \cite{Kozachenko1987sample,kraskov2004estimating} assume that
\begin{equation}
    \label{eqn:constant_density}
    P_i \approx c_{d} \ \varepsilon^{d} \ p_\mathbf{X}(\mathbf{x}_i)
\end{equation}
\noindent
where $c_{d}$ is the volume of the $d$-dimensional unit-ball ($\mathscr{B}(\varepsilon,\mathbf{x}_{i})$ with $\varepsilon = 1$). The expression for $c_{d}$ depends on the type of norm used to calculate the distances; for example, for maximum ($L_\infty$) norm $c_d = 2^d$ and for euclidean ($L_2$) norm $c_d = \pi^{d/2}/\Gamma(1 + d/2)$, where $\Gamma$ is the Gamma function. Using equation \eqref{eqn:constant_density} in equations \eqref{eqn:probability_mass_general_form} and \eqref{eqn:generic_entropy_estimate}, the entropy estimate can be written as

\begin{equation}
    \label{eqn:KL_estimate}
    \hat{H}(\mathbf{X}) =   \psi(N) - \psi(k)  +  \log(c_{d}) + \frac{d}{N} \sum_{i=1}^{N}
    \log \left(\varepsilon(i) \right)
\end{equation}
\noindent where $\varepsilon(i)$ is the distance of the $i^{th}$ sample to its $k^{th}$ nearest neighbour. This estimator is referred as KL estimator in the remainder of this article.

\subsection{The kpN estimator}
\label{sec:kpN_estimator}
Although the classical estimator works well in low-dimensions, it presents with large errors when the dimensionality of the random variable is high or the pdf in $\mathscr{B}(\varepsilon,\mathbf{x}_{i})$ shows high non-uniformity. The latter may result from: i) presence of a near-functional relationship (leading to high correlation) between two or more components of the random variable $\mathbf{X}$ \cite{gao2014Efficient}; and ii) high variability in the marginal variances of $\mathbf{X}$ in $\mathscr{B}(\varepsilon,\mathbf{x}_{i})$. In the remainder of the manuscript the term non-uniformity is used to imply the aforementioned features. The primary cause of high error in the KL estimator is the assumption of constant density in each $\mathscr{B}(\varepsilon,\mathbf{x}_{i})$. This may be unjustified when the true probability mass is likely to be high only on a small sub-region of $\mathscr{B}(\varepsilon,\mathbf{x}_{i})$. In such cases, a constant density assumption in $\mathscr{B}(\varepsilon,\mathbf{x}_{i})$ leads to and overestimation of the probability mass and hence the entropy estimate \cite{gao2014Efficient}. To remedy this, an alternate formulation for $\eta_i$ in equation \eqref{eqn:probability_mass_general_form} is sought. Contrary to a constant density assumption, the probability density in $\mathscr{B}(\varepsilon,\mathbf{x}_{i})$ is represented as
\begin{equation}
    \label{eqn:gaussian_local_approx}
    p_{\mathbf{X}}(\mathbf{x}) \approx \rho \exp \left( -\frac{1}{2}(\mathbf{x} - \boldsymbol{\mu})^T \mathbf{S}^{-1} (\mathbf{x} - \boldsymbol{\mu}) \right)
\end{equation}
\noindent
where $\boldsymbol{\mu}$ and $\mathbf{S}$ represent the empirical mean and covariance matrix of the $p$ neighbours of the point $\mathbf{x}_i$. Essentially, the probability density is assumed to be proportional to a Gaussian function approximated by using $p$-nearest neighbours of $\mathbf{x}_i$. The idea is that the $p$-neighbours would capture the local non-uniformity of the true probability density inside $\mathscr{B}(\varepsilon,\mathbf{x}_{i})$. This approach is contrary to \cite{gao2014Efficient} where the assumption of constant density is kept, and the ball is transformed using local PCA. In the proposed approach, the ball is kept constant but the probability density is assumed non-uniform. From a physical point of view, $p$ is reflective of the characteristic length of changes in the true probability distribution.

Following equation \eqref{eqn:gaussian_local_approx}, to obtain the form of equation \eqref{eqn:probability_mass_general_form}, the proportionality constant $\rho$ is obtained by requiring that the value of the local Gaussian approximation be equal to the true pdf at $\mathbf{x}_{i}$
%
%
\begin{equation}
    \label{eqn:density_gaussian_approx}
    p_{\mathbf{X}}(\mathbf{x}) \approx p_{\mathbf{X}}(\mathbf{x}_i) \ \frac{g(\mathbf{x})}{g(\mathbf{x}_i)},
\end{equation}
\noindent
where
\begin{equation}
    \label{eqn:def_gx}
    g(\mathbf{x}) = \exp \left( -\frac{1}{2}(\mathbf{x} - \boldsymbol{\mu})^T \mathbf{S}^{-1} (\mathbf{x} - \boldsymbol{\mu}) \right),
\end{equation}
\noindent
\begin{equation}
    \label{eqn:def_gxi}
  g(\mathbf{x}_i) =  \exp \left( -\frac{1}{2}(\mathbf{x}_i - \boldsymbol{\mu})^T \mathbf{S}^{-1} (\mathbf{x}_i - \boldsymbol{\mu}) \right).
\end{equation}
\noindent
Consequently, the probability mass in $\mathscr{B}(\varepsilon,\mathbf{x}_{i})$ can be written as
\begin{equation}
    \label{eqn:probability_mass_kpN}
    P_i = p_{\mathbf{X}}(\mathbf{x}_i) \ \frac{1}{g(\mathbf{x}_i)} \ G_i
\end{equation}
\noindent
where
\begin{equation}
    \label{eqn:gaussian_integral}
    G_i =
\int\displaylimits_{\mathscr{B}(\varepsilon,\mathbf{x}_{i})} g(\mathbf{x}) \ d\mathbf{x}
\end{equation}
\noindent
Using equation \eqref{eqn:probability_mass_kpN} in equations \eqref{eqn:probability_mass_general_form} and \eqref{eqn:generic_entropy_estimate}, the entropy estimate can be written as
\begin{equation}
    \label{eqn:kpN_estimator}
    \hat{H}(\mathbf{X}) =   \psi(N) - \psi(k)
    -  \frac{1}{N}\sum_{i=1}^{N} \log \left( g({\mathbf{x}_i}) \right)
    + \frac{1}{N} \sum_{i=1}^{N}
    \log  G_i .
\end{equation}

\begin{figure*}[htpb]
    \centering
    \includegraphics[width=1.0\linewidth]{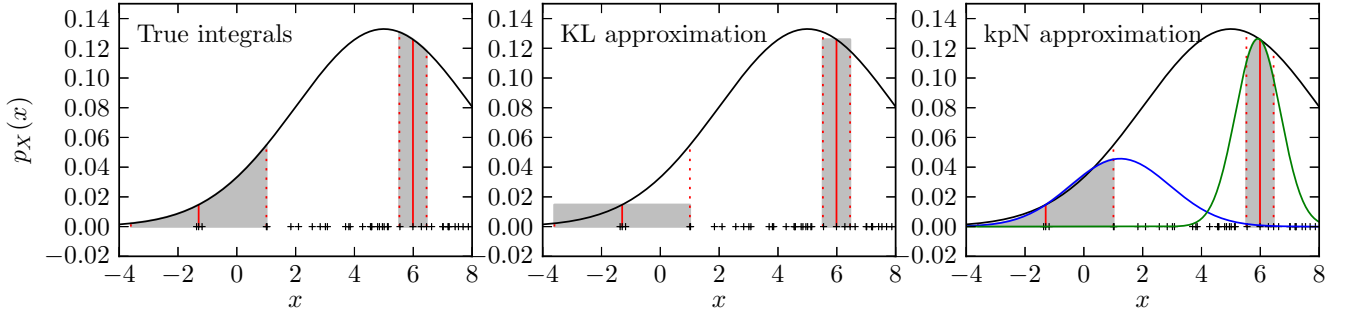}
    \caption{Demonstration of the differences between KL and kpN estimators. In each plot, the true distribution (Gaussian) is shown in solid black line and the 50 samples are shown with `+' markers. For the two points (shown in solid red vertical line), the integration region $\mathscr{B}(\varepsilon,\mathbf{x}_{i})$ with $k=3$ is shown with dashed red vertical lines, and  the integrals are shown in shaded grey. In the left panel, the true area of integration is shown. The centre panel shows the KL approximation to this area, and the right panel shows the area approximations by the kpN estimator with $p=10$. The local Gaussian approximations for the kpN estimator are shown in blue and green.}
    \label{fig:demo}
\end{figure*}

\noindent
The above estimator for entropy is referred as the kpN estimator. In this estimator, while the evaulation of $g(\mathbf{x}_i)$ is straightforward, the evaluation of $G_i$ in equation \eqref{eqn:gaussian_integral} for each sample point is not trivial, especially in high dimensions.
Before describing a computationally efficient method to evaluate this integral in the next section, a graphical demonstration of the difference in the integrals of probability density considered by the KL and kpN estimators is shown in Figure \ref{fig:demo}. Two different points -- one near the tails and one near the mode -- of a Gaussian distribution are shown. While near the mode of the distribution the approximations to the integral of the probability density are similar for the two estimators, in the tails the integral is better captured by the kpN estimator as a local Gaussian is constructed. This difference, while insignificant in low dimensions can have a significant impact in higher dimensions (demonstrated in section \ref{sec:numerical_testcases}).

\subsection{Gaussian integral in boxes}
\label{sec:gaussian_integral_in_boxes}
In order to compute the function $G_i$ a multivariate Gaussian definite integral inside $\mathscr{B}(\varepsilon,\mathbf{x}_{i})$ has to be computed. Since we adopt the $L_{\infty}$ distance, this operation amounts to computing the integral of a multivariate Gaussian inside a box. Among the methods proposed in the literature (see for instance \cite{genz1992numerical}), the Expectation Propagation Multivariate Gaussian Probability (EPMGP) method, proposed in \cite{CunninghamHL2012}, is chosen. The method is based on the introduction of a fictitious probability distribution, whose Kullback-Leibler distance with respect to the original distribution is minimised, inside the box. Since the minimisation of the Kullback-Leibler distance is equivalent, for the present setting, to the moment matching, the zero-th, first and second moments of the fictitious distribution match the ones of the original distribution. The zero-th order moment, in particular, is the sought integral value.
This method, as shown in \cite{CunninghamHL2012}, is precise in computing the definite Gaussian integral when the domain is a box.\\

Algorithm \ref{alg:the_algorithm} shows the steps to obtain the kpN estimate.

\SetInd{0.25em}{1em}
\begin{algorithm}[tb]
\small
 \KwIn{
\begin{itemize}
  \setlength{\itemsep}{4pt}
  \setlength{\parskip}{0pt}
  \setlength{\parsep}{0pt}
\item $\mathbf{x}_i \in \mathbb{R}^{d}$, $i=1\ldots N$: the samples
    \item $k$: the number of nearest neighbours for calculating $\mathscr{B}(\varepsilon,\mathbf{x}_{i})$
    \item $p$: the number of nearest neighbours for calculating the local Gaussian approximation ($p\geq k$)
\end{itemize}
 }
 \KwOut{
     $\hat{H}(\mathbf{X})$: the kpN entropy estimate
 }
\BlankLine
\BlankLine

\For{i $\leftarrow$ $\mathrm{1}$ \KwTo $N$}{ $\{\mathbf{x}_i\}^p$ $\leftarrow$ set of $p$-nearest neighbours of $\mathbf{x}_i$  ($L_\infty$ norm)
}

\BlankLine

$\hat{H}(\mathbf{X}) =   \psi(N) - \psi(k)$

\For{i $\leftarrow$ $\mathrm{1}$ \KwTo $N$}{
    $\varepsilon_i$ $\leftarrow$ $L_\infty$ distance to the $k$-th nearest neighbour of $\mathbf{x}_i$

    $\mathscr{B}(\varepsilon,\mathbf{x}_{i})$ $\leftarrow$ $\mathbf{x}_i \pm \varepsilon_i \; \mathbf{e}$ ; $\mathbf{e}$ being the canonical basis

$\mu_i$ $\leftarrow$ mean of $\{\mathbf{x}_i\}^p$

$\mathbf{S}_i$ $\leftarrow$ covariance of $\{\mathbf{x}_i\}^p$

$G_i$ $\leftarrow$ integral in equation \eqref{eqn:gaussian_integral} through EMPGP of $\mu_i$ and $\mathbf{S}_i$  (section \ref{sec:gaussian_integral_in_boxes})

$g(\mathbf{x}_i)$ $\leftarrow$ equation \eqref{eqn:def_gxi}

$\hat{H}(\mathbf{X})$ $\leftarrow$  $ \hat{H}(\mathbf{X}) + N^{-1} \left [\log(G_i) - \log(g(\mathbf{x}_i)) \right]  $
}
\caption{Algorithm to estimate kpN entropy}
\label{alg:the_algorithm}
\end{algorithm}

\section{Heuristics on the error}
\label{sec:heuristics}
In this section analytical heuristics on the error are presented to motivate the approach proposed in this work. First, the error of the KL estimator is derived. The result shows that the estimate is sensitive to both the space dimension and non-uniformity of the pdf in $\mathscr{B}(\varepsilon,\mathbf{x}_{i})$.

In what follows,  $\mathscr{B}(\varepsilon,\mathbf{x}_{i}) =[\mathbf{x}_i - \varepsilon_i,\mathbf{x}_i+ \varepsilon_i]^d$. Let $p_i(\boldsymbol{\xi})$ be the probability density in $\mathscr{B}(\varepsilon,\mathbf{x}_{i})$. In each ball, it is supposed to be $p_i(\boldsymbol{\xi})\in C^2(\mathscr{B}(\varepsilon,\mathbf{x}_{i}))$. Albeit quite strong, this regularity is introduced for sake of simplicity of the heuristics. The probability mass is $P_i = \int_{\mathscr{B}(\varepsilon,\mathbf{x}_{i})} p_i \ d\boldsymbol\xi$.

\subsection{KL estimator error analysis}
The error of the KL estimator is analysed. It comprises of two contributions: a statistical error related to the MC integration and an analytical error, resulting from the hypothesis of constant density in $\mathscr{B}(\varepsilon,\mathbf{x}_{i})$.

\subsubsection{Error in the approximation of probability mass}
The analytical contribution to the error is analysed in this section (see details in Appendix).
By considering a second order Taylor expansion of the pdf in $\mathscr{B}(\varepsilon,\mathbf{x}_{i})$, the probability mass can be approximated by:
\begin{equation}
    P_i \approx P_i^\mathrm{(KL)} + \frac{1}{2} \int_{\mathscr{B}(\varepsilon,\mathbf{x}_{i})}(\boldsymbol{\xi}-\mathbf{x}_i)^T H_{\mathbf{x}_i} (\boldsymbol{\xi} - \mathbf{x}_i)\ d\boldsymbol\xi,
\end{equation}
where $P_i^\mathrm{(KL)}$ is the probability mass resulting from constant density assumption in the KL estimator, and $H_{\mathbf{x}_i}$ is the Hessian of the pdf computed at $\mathbf{x}_i$.

Let the error in the approximation of $P_i$ be $e_{P_i}^{\mathrm{(KL)}}:=|P_i - P_i^{\mathrm{(KL)}}|$. Then:
\begin{equation}
\frac{ |\lambda^{min}_i|}{3}d 2^{d-1}\varepsilon_i^{d+2} \leq e_{P_i}^{\mathrm{(KL)}} \leq \frac{ |\lambda^{max}_i|}{3}d 2^{d-1}\varepsilon_i^{d+2},
\label{eq:e_P_bound}
\end{equation}
where $\lambda^{min,max}$ denote respectively the minimum and maximum eigenvalues of the Hessian. The lower bound can thus vanish. Concerning the upper bound, note the dependence on the dimension $d$ as well as on the maximum eigenvalue, which can be very large in the presence of non-uniformity of the pdf.

\subsubsection{Error in entropy estimation}
Let $H^{\mathrm{(KL)}}$ denote the KL entropy estimate. After some derivation and by introducing the approximation of the KL estimator in the ball, it holds:
\begin{align}
\psi(k) - \psi(N) = \frac{1}{N}\sum_i^N \log (p_i) + \frac{d}{N}\sum_i^N \log (2\varepsilon_i) + \notag\\ \frac{1}{N} \sum_i^N\log \left(1+\frac{h_i}{P_i^{\mathrm{(KL)}}} \right),
\end{align}
\noindent
where $h_i = \frac{1}{2} \int_{\mathscr{B}(\varepsilon,\mathbf{x}_{i})}(\boldsymbol{\xi}-\mathbf{x}_i)^T H_{\mathbf{x}_i} (\boldsymbol{\xi} - \mathbf{x}_i)\ d\boldsymbol\xi$.
After some algebra, the following expression for the entropy estimation is obtained:
\begin{equation}
H - H^{\mathrm{(KL)}} = e_S + \frac{1}{N} \sum_i^N \log \left(1+\frac{h_i}{P_i^{\mathrm{(KL)}}} \right),
\end{equation}
where $e_S$ is the statistical error due to the MC approximation, and the last term on the right hand side is the analytical error.

Eq.\eqref{eq:e_P_bound} and the standard $\log$-inequality (see Appendix) allows to state  the upper and lower bounds for the error:
\begin{align}
| H - H^{\mathrm{(KL)}} |  \leq e_S + \frac{d\ 2^{d-1}}{3N}\sum_i^N \frac{|\lambda_i^{max}|}{P_i^{\mathrm{(KL)}}} \varepsilon_i^{d+2}.
\label{eq:KL_upper_bound}
\end{align}
\begin{align}
| H - H^{\mathrm{(KL)}} |  \geq \left | e_S + \frac{d\ 2^{d-1}}{3N}\sum_i^N \frac{\lambda_i^{min}\; \varepsilon_i^{d+2}}{P_i^{\mathrm{(KL)}} +  \frac{| \lambda^{max}_i| d 2^{d-1}}{3} \varepsilon_i^{d+2} }  \right|
\label{eq:KL_lower_bound}
\end{align}
The error is thus bounded by the statistical error and an analytical contribution. If the distribution is piecewise linear, then the analytical contribution vanishes ($\lambda^{max}_i = 0, \forall i$ in Eq.\eqref{eq:KL_upper_bound}) since the Hessian vanishes. This corresponds to a particular case that hardly represents realistic probability distributions. The lower bound Eq.\eqref{eq:KL_lower_bound} can vanish for particular distributions. The analysis of the expressions reveals that, given a target distribution, the error in the entropy estimate can be significant in the presence of non-uniformity (high $\lambda^{max}$), and when the dimension ($d$) is high.

\subsection{The kpN estimator error analysis}
The analysis presented for the KL estimator is repeated in this section for the kpN estimator. The error analysis shows that the choice made allows to keep the structure of the KL estimator while mitigating the analytical contribution to the error. The main difference is in the approximation of the probability mass.

\subsubsection{Error in the approximation of the probability mass}
The details of the computation are presented in the Appendix. The main difference with respect to the KL estimator consists in the fact that, by constructing a Gaussian osculatory interpolant (empirically identified by using $p-$neighbours), an approximation of the Hessian of the distribution  is obtained. This estimate can be rough, but is beneficial in two cases: when the probability distributions are in a high dimensional space, or the pdf in $\mathscr{B}(\varepsilon,\mathbf{x}_{i})$ exhibits non-uniformity.

The probability mass approximation in the kpN estimator is denoted by $P_i^{(G)}$ and it is defined as:
\begin{equation}
P_i^{(G)} = P_i^{\mathrm{(KL)}} + \frac{p(\mathbf{x}_i)}{2 g(\mathbf{x}_i)} \int_{\mathscr{B}(\varepsilon,\mathbf{x}_{i})} (\boldsymbol{\xi}-\mathbf{x}_i)^T \left[\nabla \nabla g |_{\mathbf{x}_i} \right](\boldsymbol{\xi}-\mathbf{x}_i) \ d\boldsymbol\xi,
\end{equation}
so that is it the sum of the probability mass of the KL estimator and a term that approximate the Hessian of the distribution. The error estimate is:
\begin{align}
e_{P_i}^{(G)} = \frac{1}{2} \int_{\mathscr{B}(\varepsilon,\mathbf{x}_{i})}(\boldsymbol{\xi}-\mathbf{x}_i)^T \left[\nabla \nabla R |_{\mathbf{x}_i} \right] (\boldsymbol{\xi}-\mathbf{x}_i) \ d\boldsymbol\xi,
\end{align}
where $R$ is the difference between the target distribution and its gaussian approximation inside the box.

\subsubsection{Error in the approximation of the entropy}
By repeating the same analysis as for the KL estimator, the following upper and lower bounds are obtained:
\begin{align}
| H - H^{(G)} |  \geq \left | e_S + \frac{d\ 2^{d-1}}{3N}\sum_i^N \frac{\zeta_i^{min}\; \varepsilon_i^{d+2}}{P_i^{\mathrm{(KL)}} +  \frac{| \zeta^{max}_i| d 2^{d-1}}{3} \varepsilon_i^{d+2} }  \right|,
\label{eq:G_lower_boundText}
\end{align}
\begin{align}
| H - H^{(G)} |  \leq e_S + \frac{d\ 2^{d-1}}{3N}\sum_i^N \frac{|\zeta_i^{max}|}{P_i^{\mathrm{(KL)}}} \varepsilon_i^{d+2},
\label{eq:G_upper_boundText}
\end{align}
where $\zeta^{min,max}$ are the maximum and minimum eigenvalues of the Hessian of the residual $R$.

Let us remark that the behaviour is the same for the KL estimator and the kpN estimator in terms of functional dependence with respect to the space dimension. However, by approximating the Hessian (avoiding a bad choice of $k$ and $p$ is important to this end), $\zeta_i^{max}$ can be significantly lower than $\lambda_i^{max}$. This has two potential advantages: first, that in the presence of non-uniformity, the upper bound on the kpN error is smaller; and second that, even if correlations are not significant, a lower $\zeta_i^{max}$ results in a lower rate of increase of error with increasing dimensions. \\

\section{Numerical testcases}
\label{sec:numerical_testcases}
In this section, the numerical experiments are presented. The first test case aims at validating the proposed approach against analytical results, in simple settings.

Then, several relevant properties of the methods are investigated in more complicated settings, that frequently occur when realistic datasets are considered. First, the robustness in dimension increase is investigated. Then, the entropy estimation in presence of functional dependency leading to high correlation is shown.

\begin{table*}[htpb]
    \setlength{\tabcolsep}{0.3cm}
    \centering
    \caption{Summary of the distributions for the analysis of $k$, $p$, and $N$}
    \label{tab:kpn_distributions}
    \begin{tabular}{|>{\centering\arraybackslash}p{6cm}|c|}
        \hline
        \textbf{ Distribution } & \textbf{Parameters} \\[4pt]
        \hline
        \parbox{6cm}{2-D Multivariate Normal \\ (correlation coefficient $r$)} &
        \begin{tabular}{ccc}
            mean           & variance   & $r$ \\
            $[0.0, 0.0]$ & $[1.0, 1.0]$ & 0.5 \\
        \end{tabular} \\[4pt]
        \hline
        \parbox{6cm}{3-D Gamma distribution \\ (Independent along each dimension)} &
        \begin{tabular}{cccccc}
        $k_1$   & $\theta_1$ & $k_2$ & $\theta_2$ & $k_3$ & $\theta_3$ \\
            1.5 & 2.0        & 3.0   & 2.5        & 20.0  & 1.0 \\
        \end{tabular} \\[4pt]
        \hline
        \parbox{6cm}{4-D Beta distribution \\ (Independent along each dimension)} &
        \begin{tabular}{cccccccc}
            $\alpha_1$ & $\beta_1$ & $\alpha_2$ & $\beta_2$ & $\alpha_3$ & $\beta_3$ & $\alpha_4$ & $\beta_4$ \\
            2.0        & 2.0       & 2.0        & 5.0       & 0.5        & 0.5       & 5.0        & 1.0 \\
        \end{tabular} \\
        \hline
    \end{tabular}
\end{table*}

\subsection{Analysis of estimator: effect of $k$, $p$, and $N$ }
\label{sec:kpn_analysis}
\begin{figure*}[htbp]
    \subfloat[\% relative error in the entropy estimate]{%
      \includegraphics[width=1.0\textwidth]{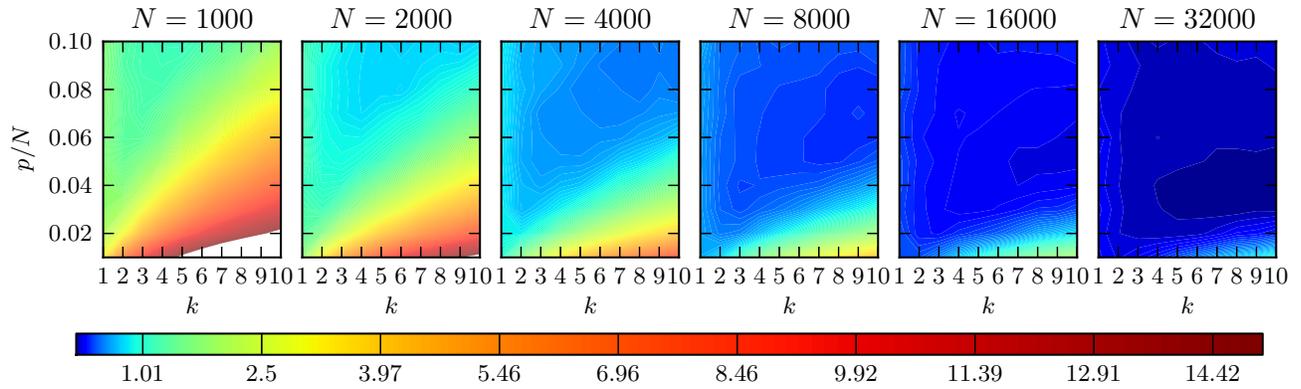}
    \label{fig:Normal2D_mean}
    }
    \hfill
    \subfloat[\% variance of relative error in the entropy estimate]{%
      \includegraphics[width=1.0\textwidth]{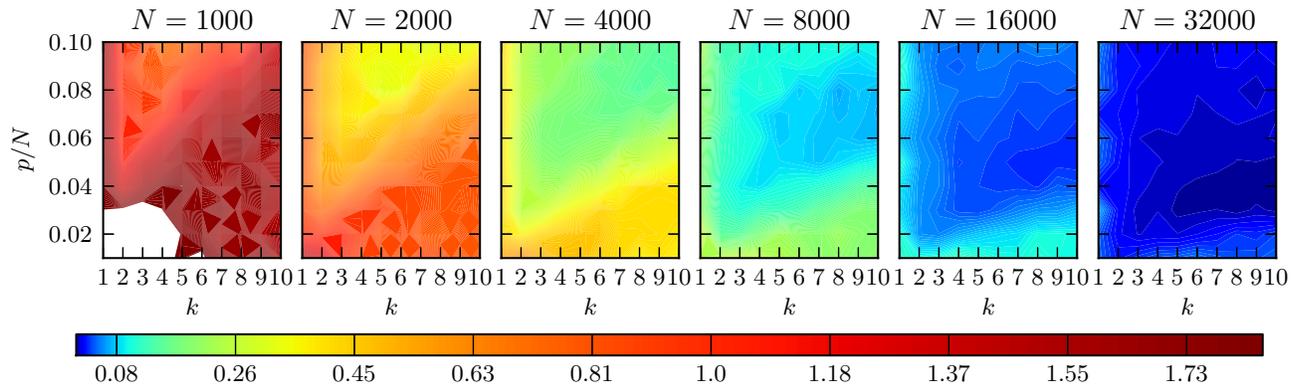}
      \label{fig:Normal2D_var}
    }
    \caption{kpN entropy estimate for 2D-Gaussian distribution with correlation $r=0.5$ (see Table \ref{tab:kpn_distributions})}
    \label{fig:NkpAnalysis_Normal}
  \end{figure*}

\begin{figure*}[htbp]
    \subfloat[\% relative error in the entropy estimate]{%
      \includegraphics[width=1.0\textwidth]{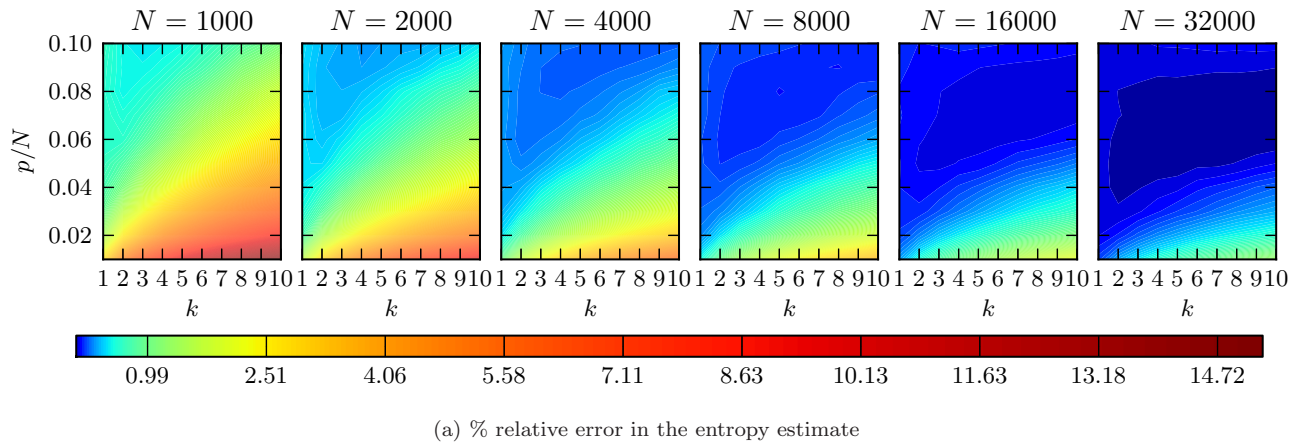}
    \label{fig:Gamma3D_mean}
    }
    \hfill
    \subfloat[\% variance of relative error in the entropy estimate]{%
      \includegraphics[width=1.0\textwidth]{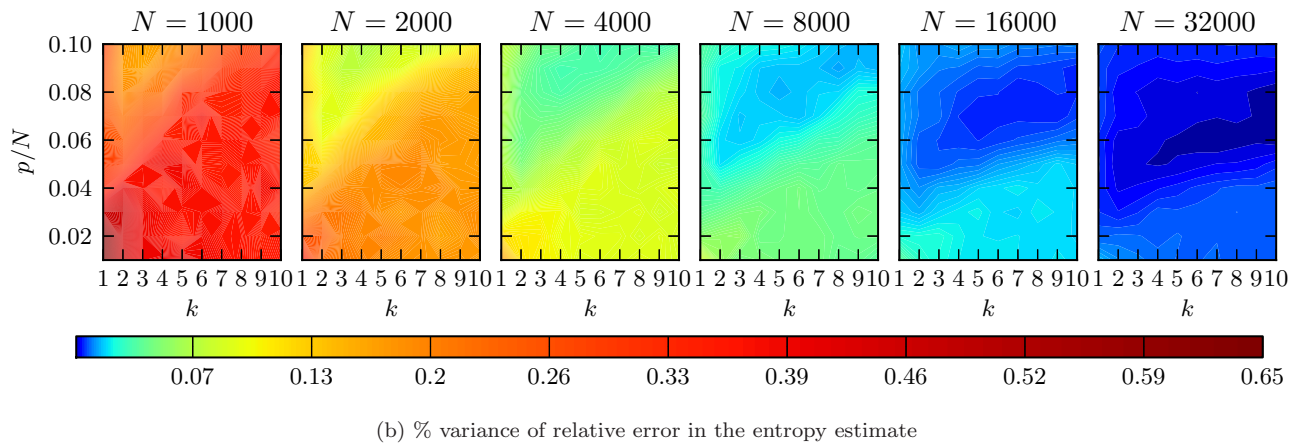}
      \label{fig:Gamma3D_var}
    }
    \caption{kpN entropy estimate for 3D-Gamma distribution with shape parameters shown in Table \ref{tab:kpn_distributions}}
    \label{fig:NkpAnalysis_Gamma}
  \end{figure*}

\begin{figure*}[htbp]
    \subfloat[\% relative error in the entropy estimate]{%
      \includegraphics[width=1.0\textwidth]{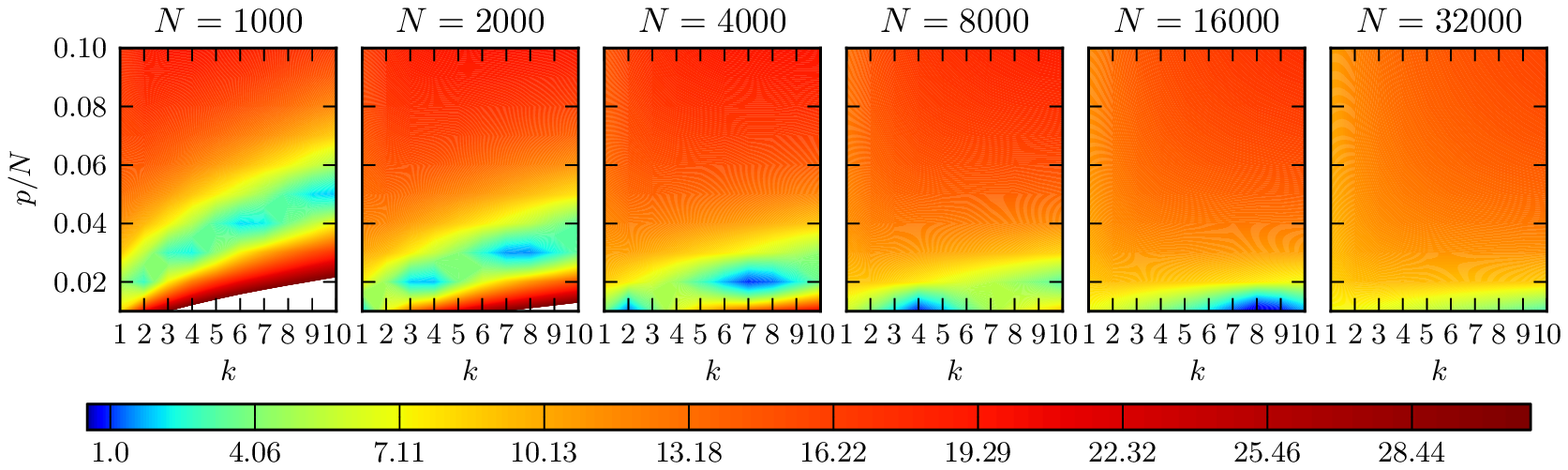}
    \label{fig:Beta4D_mean}
    }
    \hfill
    \subfloat[\% variance of relative error in the entropy estimate]{%
      \includegraphics[width=1.0\textwidth]{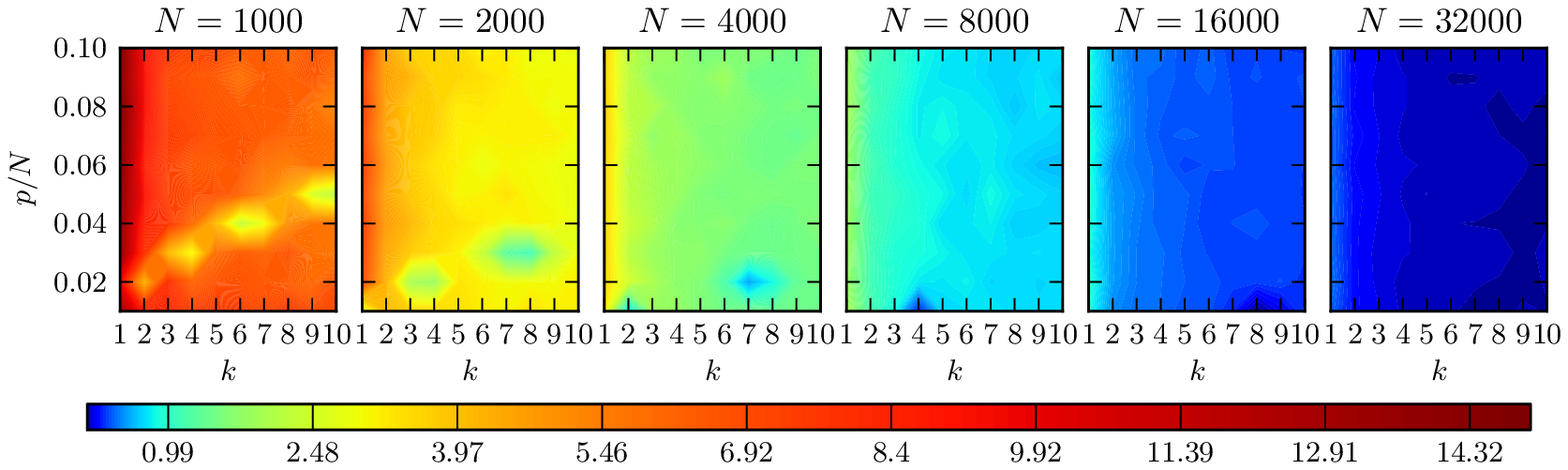}
      \label{fig:Beta4D_var}
    }
    \caption{kpN entropy estimate for 4D-Beta distribution with shape parameters shown in Table \ref{tab:kpn_distributions}}
    \label{fig:NkpAnalysis_Beta}
  \end{figure*}

To assess the effect of the parameters $k$, $p$, and $N$, in the kpN estimator, three probability distributions in two, three, and four dimensions are considered. A summary of the these distributions is presented in Table \ref{tab:kpn_distributions}. For all the three distributions, the number of samples $N$ are varied from 1000 to 32000, $k$ is varied from 1 to 10, and $p/N$ is varied from 0.01 to 0.10. For each set of these parameters an $N_\mathrm{ens}=1000$ independent kpN entropy estimates are calculated and the corresponding mean and variance of the error with respect to the analytically known true entropy is calculated. These results for the 2-D Gaussian, 3-D Gamma, and 4-D Beta distributions are shown in Figures \ref{fig:NkpAnalysis_Normal}, \ref{fig:NkpAnalysis_Gamma}, and \ref{fig:NkpAnalysis_Beta}, respectively. From these plots it is observed that the variance of the error decreases with increasing $N$ as expected. Furthermore, the variance appears to be high for $k=1,2$ and then lower and approximately invariant with increasing $k$. This is consistent with the behaviour of the KL estimator \cite{kraskov2004estimating}.
Recall that the parameter $p$ is reflective of the length-scale of changes in the probability density. For a Gaussian distribution, it is clear that a higher $p$ will result in lower error as the local Gaussian approximations will better approximate the true distribution. This is observed in Figure \ref{fig:Normal2D_mean}. A similar behaviour is observed for the Gamma distribution (Figure \ref{fig:Gamma3D_mean}), but for the Beta distribution (Figure \ref{fig:Beta4D_mean}) a clear optimal range of $p/N$ varying from 0.01 to 0.05 can be identified. The length-scale of the density variation will in general not be known \emph{a priori} (especially in higher dimensions where the samples are hard to visualise) and consequently a large $p/N$ should be avoided. From Figures \ref{fig:Normal2D_mean}, \ref{fig:Gamma3D_mean}, and \ref{fig:Beta4D_mean}, it is observed that unless a particularly bad combination of the $N$, $k$, and $p$, parameters -- specifically low $N$, high $k$, and small $p/N$ -- is chosen, the errors across the entire spectrum of parameter variations are less than 10\%. Overall, based on the performance of the estimator across the three significantly different distributions considered, $k$ is recommended to be chosen between 3 and 5, and $p/N$ between 0.02 and 0.05.

\subsection{Dimension increase}
The properties of the kpN estimator regarding robustness to dimension increase are investigated. For all these tests $N=10000$, $k=4$, $p/N = 0.02$ are fixed. The method is compared to the standard KL estimator in multi-dimensional uncorrelated Gaussian, Gamma and Beta distributions. The dimension ranges from $4$ to $80$. For all the cases, the quantity of interest is the relative error, defined as $e = \frac{|H^* - H |}{H^*}$, where $H^*$ is the analytical value. The distributions used to test the method are quite regular and smooth. Moreover, no correlation is considered, the only focus being the behaviour with respect to the dimension increase. For all the tests, the computations were repeated $N_\mathrm{ens} = 1000$ times, and corresponding mean-values and variances are reported.

\subsubsection{Multi-dimensional Gaussian}
The first test case is the entropy computation of a multi-dimensional Gaussian:
\begin{equation}
p^* = \prod_i^d g_i(\mu_i,\sigma_i),
\end{equation}
where $d$ is the space dimension, $\mu_i=0, \ \forall i$. The variance $\sigma_i$ ranges uniformly in $[0.2,2]$, \emph{i.e.} $\sigma_i = 1.8 (i-1)/(d-1) + 0.2$.
\begin{figure}
\includegraphics[height=5.0cm, width=0.5\textwidth]{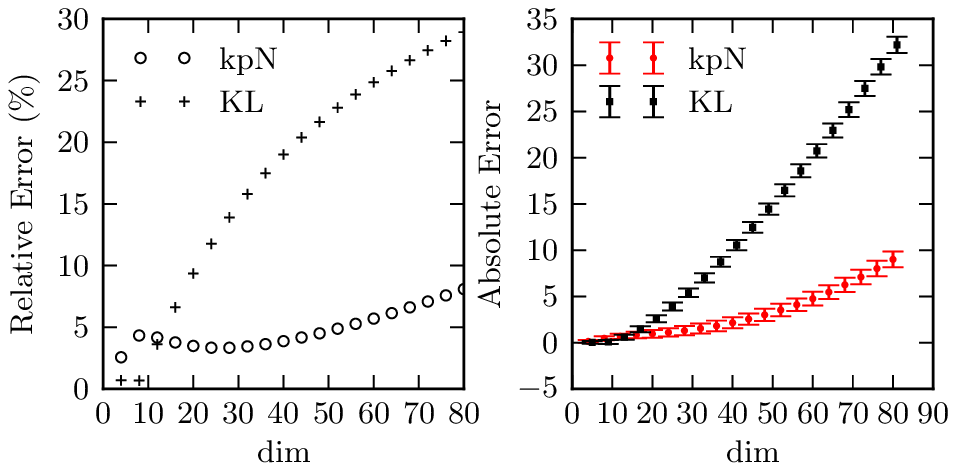}
\caption{Error analysis of a multivariate gaussian}
\label{fig:dimAnalysisNormal}
\end{figure}
The results are summarised in Fig.\ref{fig:dimAnalysisNormal}. The relative error at low dimension is higher than that of the KL estimator. This is due to the fact that the parameters adopted are not optimal for this distribution, given the number of sample (a higher value of $p/N$ would provide a better result). The kpN error is significantly smaller when the dimension increases: namely, at dimension $d=80$, it has an error which is less than $10\%$, while the KL estimator has an error which is about three times larger, despite the fact that the probability distribution is quite regular.

\subsubsection{Multi-dimensional Gamma}
The case of a multivariate Gamma distribution is commented. Similarly to earlier case, the distribution is defined as a product of univariate distributions:
\begin{equation}
p^* = \prod_i^d \gamma_i(k_i,\theta_i),
\end{equation}
where $k_i$ and $\theta_i$ are the shape and scale parameters of the distribution. The shape parameter $k_i$ varies uniformly in $[0.5, 5.0]$ while the scale parameter $\theta_i$ varies in $[1.0,2.0]$.
\begin{figure}
\includegraphics[height=5.0cm, width=0.5\textwidth]{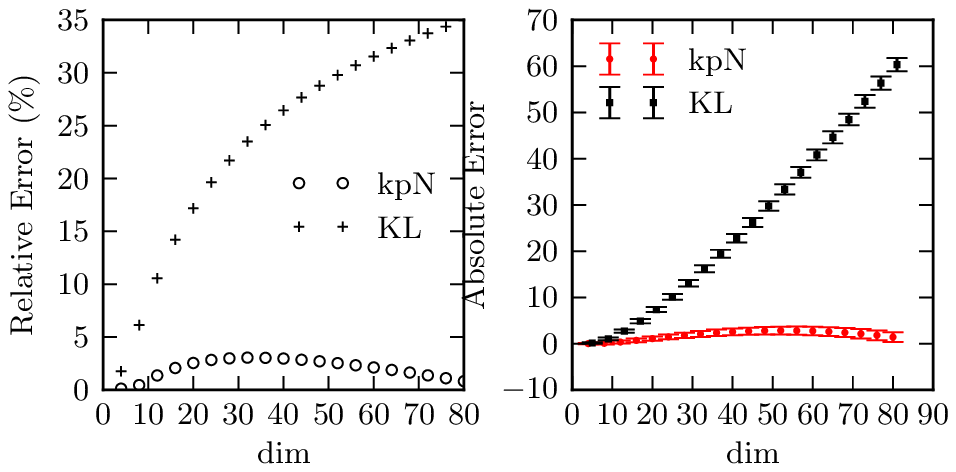}
\caption{Error analysis of a multivariate gamma}
\label{fig:dimAnalysisGamma}
\end{figure}
The results are shown in Fig.\ref{fig:dimAnalysisGamma}. For this case, the kpN estimator always outperforms the KL estimator. Note that the error is not necessarily monotonic with respect to the dimension of the space. This depends on the particular nature of the distribution as well as on the parameters $k$ and $p$ adopted. Nonetheless, the kpN error is less than $5\%$ across the entire range of dimensions considered, while the KL error grows up to $35\%$.

\subsubsection{Multi-dimensional Beta}
The last test case shown concerns the entropy estimation for a multivariate Beta distribution of the form:
\begin{equation}
p^* = \prod_i^d \beta_i(\alpha_i,\beta_i),
\end{equation}
where $\alpha_i$ varies in $[0.5,5.0]$ and $\beta_i$ varies in $[0.5,5.0]$.
\begin{figure}
\includegraphics[height=5.0cm, width=0.5\textwidth]{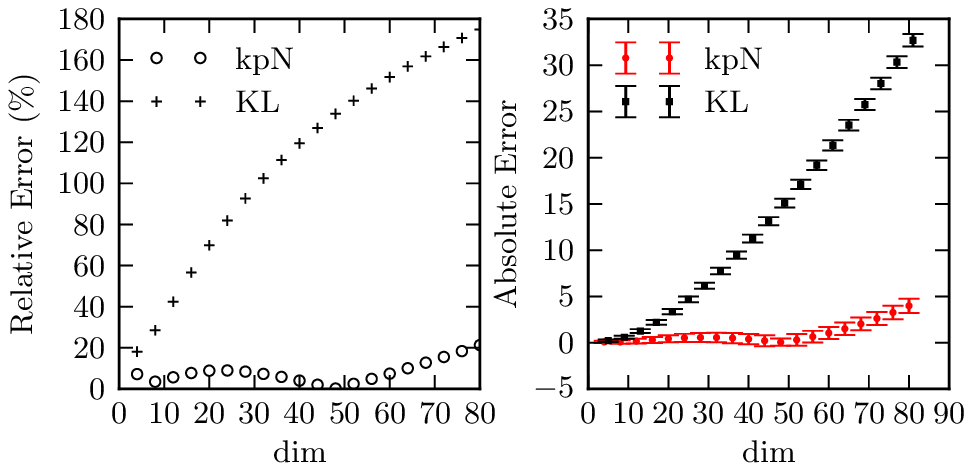}
\caption{Error analysis of a multivariate beta}
\label{fig:dimAnalysisBeta}
\end{figure}
The results of the kpN and KL entropy estimates are shown in Figure \ref{fig:dimAnalysisBeta}. This test appears to be most critical as, on average, the errors on both KL and kpN estimates are higher when compared to the previous Gaussian and Gamma distributions. This may partly be due to pathological nature of the Beta distribution for particular choices of the $\alpha$ and $\beta$ parameters (for example $\alpha = \beta = 0.5$), or (although unclear why) due to the fact that the Beta distribution has only a finite support over [0.0,1.0] in all dimensions. From Figure \ref{fig:dimAnalysisBeta}, the error of the kpN estimator is always less than $20\%$ whereas the KL estimate has a relative error of about $150\%$, which is almost one order of magnitude higher.

\subsubsection{Discussion}
The three tests presented aim at investigating the behaviour of the estimator with respect to the dimension of the space. The error of the KL estimator is monotonic and grows quite fast, because the analytical contribution to the error grows significantly with the dimension, when the number of sample is kept fixed. On the contrary, the kpN estimator proposed manages to mitigate this error by providing a rough estimate of the Hessian of the distribution in each box. The proposed kpN estimator is more robust to the dimension increase, or, conversely, given a certain dimension of the space, it allows to estimate the entropy by using a smaller number of samples. This feature is particularly appealing when dealing with the analysis of realistic datasets.

\subsection{Functional dependency and correlation}
\label{sec:corr}
Another interesting aspect that occurs frequently when realistic applications are considered is the possible presence of correlation. In this section, the robustness of the entropy estimators is investigated: the kpN method is compared to the KL method for fixed parameters: $N=5000$, $k=4$, $p/N=0.02$. A simple test case is proposed: the entropy of a Gaussian distribution on a linear manifold is computed, with different levels of noise. The system is
\begin{equation}
y = t x + \nu,
\end{equation}
where $x$ is a normal random variable with zero mean and unit variance, $t \in \mathbb{R}^+$ is a positive scalar, and $\nu$ is a normal random variable with zero mean and variance $\sigma_n^2$.
\begin{figure}
\includegraphics[height=5.0cm, width=0.5\textwidth]{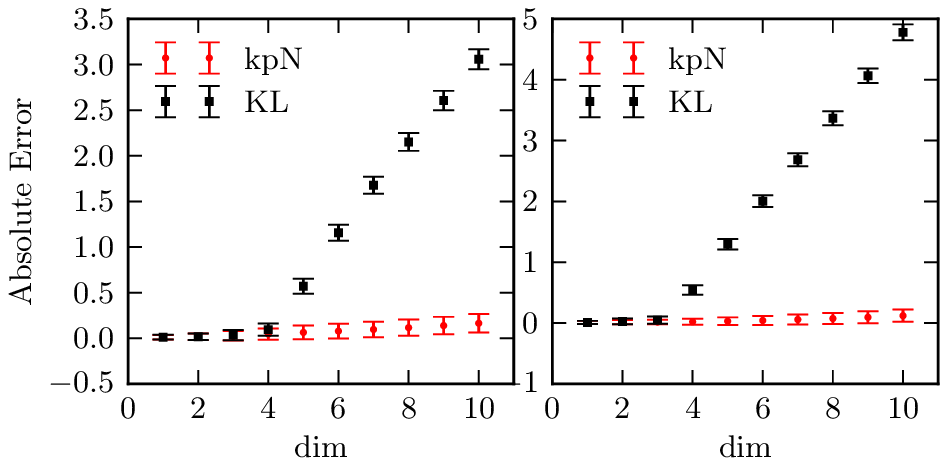}
\caption{Error analysis for a gaussian on a linear manifold. Absolute error with respect to the dimension, level of noise: $\sigma^2_n = 10^{-1}$ (left) and $\sigma^2_n = 10^{-3}$ (right) }
\label{fig:linMan}
\end{figure}

The system output $y$ is observed at discrete times $t_i = \left\lbrace 1,\ldots,9 \right\rbrace$, providing $y_i = y(t_i)$. The objective is to estimate the entropy of the joint probability distribution of $[x, y_1, \ldots, y_i]$ for increasing $i$. For this test case, two different levels of noise are considered, namely $\sigma_n^2 = \{10^{-1}, 10^{-3}\}$. The joint dimension increases up to $d=10$. The results (in terms of absolute error and variance) are shown in Fig.\ref{fig:linMan} for $\sigma_n^2 = 10^{-1}$ and $\sigma_n^2 = 10^{-3}$. When the dimension is low, the performances of the KL estimator and that of the proposed kpN estimator are comparable, \emph{i.e.} no significant difference in error is observed in terms of both the means and the variances. When the dimension increases, depending on the level of noise, the KL estimator starts deviating from the true estimate, whereas the proposed kpN estimator provides a significantly better result. The higher the noise level, the better is the behaviour of the classical KL estimator. This apparently paradoxical result can be explained by considering the analytical heuristics proposed. When the level of noise is higher, the samples are less correlated and, thus, the maximum eigenvalue of the Hessian is, on average, smaller. The joint distribution being more regular, a better entropy estimate is obtained by the classical KL estimator. The kpN estimator, on the other hand, is more robust to variations in noise-levels as based on the $p$-neighbours the covariance of the local Gaussian approximation adjusts accordingly.

\section{Conclusions and Perspectives}
A new $k$-nearest neighbour based entropy estimator, that is efficient in high dimensions and in the presence of large non-uniformity, is proposed. The proposed idea relies on the introduction of a Gaussian osculatory interpolation, which in-turn is based on an empirical evaluation of $p$-nearest neighbours. By this introduction, the local non-uniformity of the underlying probability distribution is captured, while retaining all the appealing computational advantages of classical kNN estimators. The robustness of the new estimator is tested for a variety of distributions -- ranging from infinite support Gamma distributions to finite support Beta distributions -- in successively increasing dimensions (up to 80). Furthermore, a case of direct functional relationship leading to high correlations between the components of a random variable is considered. Across all the tests, the new estimator is shown to consistently outperform the classical kNN estimator.

The main perspective of the current work is that the proposed estimator can be used as a building block to construct estimators for other quantities of interest such as mutual information, particularly in high dimensions. Another perspective is the development of strategies to automatically adapt $p$ based on properties of the cloud of local samples.

\section{Appendix}
\label{sec:App_A}
In this Appendix, the details of the error heuristics are presented.
Let us recall the notation and the main hypotheses. The $\varepsilon$-ball is denoted by $\mathscr{B}(\varepsilon,\mathbf{x}_{i})=[\mathbf{x}_i - \varepsilon_i,\mathbf{x}_i+ \varepsilon_i]^d$. Let $p_i(\boldsymbol\xi)\in C^2(\mathscr{B}(\varepsilon,\mathbf{x}_{i}))$ be the probability density in $\mathscr{B}(\varepsilon,\mathbf{x}_{i})$. The probability mass in $\mathscr{B}(\varepsilon,\mathbf{x}_{i})$ is $P_i = \int_{\mathscr{B}(\varepsilon,\mathbf{x}_{i})} p_i \ d\boldsymbol\xi$.

\subsection{KL estimator error analysis}
The error of the KL estimator is analysed. First, the error on the probability mass in a generic $\mathscr{B}(\varepsilon,\mathbf{x}_{i})$ is computed, and the result is used to compute the error on the entropy.

\subsubsection{Error in the approximation of the probability mass}
The analytical contribution to the error is due to the approximation of the probability mass $P_i$. Consider a Taylor expansion of $p_i$ centered around $x_i$:
\begin{align}
P_i = \int_{\mathscr{B}(\varepsilon,\mathbf{x}_{i})} p(\mathbf{x}_i) + (\boldsymbol{\xi} - \mathbf{x}_i)\cdot\nabla p|_{\mathbf{x}_i} + \notag\\ \frac{1}{2} (\boldsymbol{\xi}-\mathbf{x}_i)^T H_{\mathbf{x}_i} (\boldsymbol{\xi} - \mathbf{x}_i) + o(|\boldsymbol{\xi}-\mathbf{x}_i|^2) \ d\boldsymbol\xi,
\end{align}
where $H_{\mathbf{x}_i}$ is the Hessian computed in $x_i$. The first term of the series yields the KL approximation $P_i^{\mathrm{(KL)}}$, the second term vanishes since it is the integral of an even function over a symetric interval, the third term represent the error of the approximation:
\begin{equation}
P_i \approx P_i^{\mathrm{(KL)}} + \frac{1}{2} \int_{\mathscr{B}(\varepsilon,\mathbf{x}_{i})}(\boldsymbol{\xi}-\mathbf{x}_i)^T H_{\mathbf{x}_i} (\boldsymbol{\xi} - \mathbf{x}_i)\ d\boldsymbol\xi,
\end{equation}
obtained by discarding the higher order terms. Since $P_i \geq 0$, let us make the hypothesis that this hold even for the truncated approximation, \emph{i.e.}:
\begin{equation}
\label{eq:trunc_hyp}
\frac{\left |\frac{1}{2} \int_{\mathscr{B}(\varepsilon,\mathbf{x}_{i})}(\boldsymbol{\xi}-\mathbf{x}_i)^T H_{\mathbf{x}_i} (\boldsymbol{\xi} - \mathbf{x}_i)\ d\boldsymbol\xi  \right|}{P_i^{K}} \leq 1.
\end{equation}
The integral $h_i = \frac{1}{2} \int_{\mathscr{B}(\varepsilon,\mathbf{x}_{i})}(\boldsymbol{\xi}-\mathbf{x}_i)^T H_{\mathbf{x}_i} (\boldsymbol{\xi} - \mathbf{x}_i)\ d\boldsymbol\xi$ is estimated. A standard result on the quadratic forms is used:
\begin{align}
\label{eq:quad_form}
\lambda_i^{min} |\boldsymbol{\xi}-\mathbf{x}_i|^2 \leq (\boldsymbol{\xi}-\mathbf{x}_i)^T H_{\mathbf{x}_i} (\boldsymbol{\xi} - \mathbf{x}_i) \leq \lambda_i^{max} |\boldsymbol{\xi}-\mathbf{x}_i|^2,
\end{align}
where $\lambda_i^{min,max}$ are the minimum and maximum eigenvalues of $H_{\mathbf{x}_i}$. Then, the bounds on $h_i$ are simply obtained by computing the integral over $\mathscr{B}(\varepsilon,\mathbf{x}_{i})$:
\begin{align}
\int_{\mathscr{B}(\varepsilon,\mathbf{x}_{i})} |\boldsymbol{\xi}-\mathbf{x}_i|^2 \ d\boldsymbol\xi = \sum_j^d \int_{\mathscr{B}(\varepsilon,\mathbf{x}_{i})} (\xi_j - x_{i,j})^2 \ d\boldsymbol\xi.
\end{align}
By virtue of the symetry of the ball, this integral can be computed for just one $j$ and then multiplied by $d$. Let $\mathscr{B}(\varepsilon,\mathbf{x}_{i}) = [x_{i,j}-\varepsilon_i, x_{i,j}+\varepsilon_i]\times [x_{i,k}-\varepsilon_i, x_{i,k}+\varepsilon_i]^{d-1},\ k\neq j$. It holds:
\begin{align}
\label{eq:int_over_j}
\int_{\mathscr{B}(\varepsilon,\mathbf{x}_{i})} (\xi_j - x_{i,j})^2 \ d\boldsymbol\xi = (2\varepsilon)^{d-1} \int_{-\varepsilon}^{\varepsilon} \eta^2 \ d\eta = (2\varepsilon)^{d-1} \frac{2}{3} \varepsilon^3.
\end{align}
By putting together the bounds in Eq.\eqref{eq:quad_form} and the result in Eq.\eqref{eq:int_over_j}, the error approximation is obtained. Let $e_{P_i}^{\mathrm{(KL)}}:=|P_i - P_i^{\mathrm{(KL)}}|$. Then:
\begin{equation}
\frac{ |\lambda^{min}_i|}{3}d 2^{d-1}\varepsilon_i^{d+2} \leq e_{P_i}^{\mathrm{(KL)}} \leq \frac{ |\lambda^{max}_i|}{3}d 2^{d-1}\varepsilon_i^{d+2}.
\label{eq:e_P_bound}
\end{equation}

\subsubsection{Error in the approximation of the entropy}
The error on the entropy estimate is obtained by a derivation of the KL estimator.
The KL estimator is obtained by equating $\mathbb{E} \left\lbrace \log(P) \right\rbrace = \psi(k) - \psi(N)$. Let $H^{\mathrm{(KL)}}$ denote the entropy estimated by using the KL estimator.  We write:
\begin{align}
\psi(k) - \psi(N) = \frac{1}{N}\sum_i^N \log \left(P_i^{\mathrm{(KL)}} + h_i \right).
\end{align}
The properties of the logarithm are used, leading to:
\begin{align}
\psi(k) - \psi(N) = \frac{1}{N}\sum_i^N \log \left(P_i^{\mathrm{(KL)}} \right) + \frac{1}{N} \log \left(1+\frac{h_i}{P_i^{\mathrm{(KL)}}} \right).
\end{align}
The KL approximation of $P_i^{K}$ is introduced:
\begin{align}
\psi(k) - \psi(N) = \frac{1}{N}\sum_i^N \log (p_i) + \frac{d}{N}\sum_i^N \log (2\varepsilon_i) + \notag\\ \frac{1}{N} \sum_i^N\log \left(1+\frac{h_i}{P_i^{\mathrm{(KL)}}} \right).
\end{align}
After some algebra, it holds:
\begin{equation}
\label{eq:ent_KL_err}
H - H^{K} = e_S + \frac{1}{N} \sum_i^N \log \left(1+\frac{h_i}{P_i^{\mathrm{(KL)}}} \right),
\end{equation}
where $e_S$ is the statistical error due to the MC approximation, and the last term on the right hand side is the analytical error.

The use of the result presented in Eq.\eqref{eq:e_P_bound} and of a standard $\log$-inequality allows to state upper and lower bounds for the error.

Indeed, the hypothesis in Eq.\eqref{eq:trunc_hyp} allows to make use of the following:
\begin{align}
\frac{x}{1+x} \leq \log(1+x) \leq x.
\end{align}
After having set $x = h_i / P_i^{K}$, we have:
\begin{equation}
\frac{h_i}{h_i + P_i^{\mathrm{(KL)}}} \leq \log \left(1 + \frac{h_i}{P_i^{\mathrm{(KL)}}} \right) \leq \frac{h_i}{P_i^{\mathrm{(KL)}}}.
\end{equation}
In order to get a lower bound, the left hand side is studied. It holds:
\begin{equation}
\frac{h_i}{h_i + P_i^{\mathrm{(KL)}}} \geq \frac{\min (h_i)}{\max (h_i) + P_i^{\mathrm{(KL)}}} = \frac{\lambda_i^{min} d \ 2^{d-1} \varepsilon_i^{d+2}}{ \lambda_i^{max}d \ 2^{d-1} \varepsilon_i^{d+2} + 3 P_i^{K}}.
\end{equation}
The use of this results allows to state the lower bound for the error:
\begin{align}
| H - H^{\mathrm{(KL)}} |  \geq \left | e_S + \frac{d\ 2^{d-1}}{3N}\sum_i^N \frac{\lambda_i^{min}\varepsilon_i^{d+2} }{P_i^{\mathrm{(KL)}} +  \frac{| \lambda^{max}_i| d 2^{d-1}}{3} \varepsilon_i^{d+2} } \right|.
\label{eq:KL_lower_bound}
\end{align}

In order to derive the upper bound, the right hand side of the logarithmic inequality is studied:
\begin{align}
\frac{h_i}{P_i^{\mathrm{(KL)}}} \leq \frac{\max(h_i) }{P_i^{\mathrm{(KL)}}} = \frac{\lambda_i^{max} d 2^{d-1} \varepsilon_i^{d+2}}{3 P_i^{\mathrm{(KL)}}}.
\end{align}
By using this, the upper bound reads:
\begin{align}
| H - H^{\mathrm{(KL)}} |  \leq e_S + \frac{d\ 2^{d-1}}{3N}\sum_i^N \frac{|\lambda_i^{max}|}{P_i^{\mathrm{(KL)}}} \varepsilon_i^{d+2}.
\label{eq:KL_upper_bound}
\end{align}

\subsection{Analysis of the kpN estimator}
As commented above, the main difference is in the approximation of the probability mass in $\mathscr{B}(\varepsilon,\mathbf{x}_{i})$. In particular, an osculatory interpolation with an empirically estimated multivariate Gaussian is constructed.

\subsubsection{Error in the approximation of the probability mass}
The probability density distribution inside the ball is approximated by:
\begin{equation}
p(\boldsymbol{\xi}) = p(\mathbf{x}_i)\frac{g(\boldsymbol{\xi})}{g(\mathbf{x}_i)} + R(\boldsymbol{\xi}),
\end{equation}
where $g:= \exp \left( -\frac{1}{2}(\boldsymbol{\xi} - \boldsymbol\mu)^T \mathbf{S}^{-1} (\boldsymbol{\xi} - \boldsymbol\mu) \right)$, where $\boldsymbol\mu,\mathbf{S}$ are the empirically evaluated mean and covariance, $R$ is the residual of the approximation. Since $p(\boldsymbol{\xi} = \mathbf{x}_i) = p(\mathbf{x}_i)$ by construction of the approximation, it follows $R(\mathbf{x}_i)=0$.  The Taylor expansion of the probability density distribution centred around $\mathbf{x}_i$ is computed for the gaussian approximation:
\begin{align}
p(\boldsymbol{\xi}) \approx p(\mathbf{x}_i) + (\boldsymbol{\xi}-\mathbf{x}_i)\cdot \left(\frac{p(\mathbf{x}_i)}{g(\mathbf{x}_i)}\nabla g |_{\mathbf{x}_i} + \nabla R |_{\mathbf{x}_i} \right) + \notag\\
\frac{1}{2} (\boldsymbol{\xi}-\mathbf{x}_i)^T K_{\mathbf{x}_i} (\boldsymbol{\xi}-\mathbf{x}_i),
\end{align}
where $K_{\mathbf{x}_i} = \frac{p(\mathbf{x}_i)}{g(\mathbf{x}_i)} \nabla \nabla g |_{\mathbf{x}_i} + \nabla \nabla R |_{\mathbf{x}_i} $ is the Hessian computed for the gaussian approximation.

The expression is used to compute the probability mass. Remark that, as before, the linear contribution vanishes identically due to the symetry of the ball. It holds, at second order:
\begin{align}
P_i = \int_{\mathscr{B}(\varepsilon,\mathbf{x}_{i})} p(\boldsymbol{\xi}) \ d\boldsymbol\xi \approx P_i^{\mathrm{(KL)}} + \frac{1}{2} \int_{\mathscr{B}(\varepsilon,\mathbf{x}_{i})} (\boldsymbol{\xi}-\mathbf{x}_i)^T H_{\mathbf{x}_i} (\boldsymbol{\xi}-\mathbf{x}_i) \ d\boldsymbol\xi,
\label{eq:Taylor}
\end{align}
where $H$ denotes the Hessian of the target distribution. On the other hand:
\begin{align}
P_i = \int_{\mathscr{B}(\varepsilon,\mathbf{x}_{i})} p(\boldsymbol{\xi}) \ d\boldsymbol\xi \approx P_i^{\mathrm{(KL)}} + \frac{1}{2} \int_{\mathscr{B}(\varepsilon,\mathbf{x}_{i})} (\boldsymbol{\xi}-\mathbf{x}_i)^T K_{\mathbf{x}_i} (\boldsymbol{\xi}-\mathbf{x}_i) \ d\boldsymbol\xi.
\end{align}
The expression for $K_{\mathbf{x}_i}$ is introduced, allowing to understand what the gaussian approximation does in terms of approximating the mass:
\begin{align}
\label{eq:mass_approx_g}
P_i = P_i^{\mathrm{(KL)}} + \frac{1}{2} \int_{\mathscr{B}(\varepsilon,\mathbf{x}_{i})} (\boldsymbol{\xi}-\mathbf{x}_i)^T \left[ \frac{p(x_i)}{g(x_i)} \nabla \nabla g |_{\mathbf{x}_i} \right](\boldsymbol{\xi}-\mathbf{x}_i) \ d\boldsymbol\xi + \notag\\
 + \frac{1}{2} \int_{\mathscr{B}(\varepsilon,\mathbf{x}_{i})}(\boldsymbol{\xi}-\mathbf{x}_i)^T \left[\nabla \nabla R |_{\mathbf{x}_i} \right] (\boldsymbol{\xi}-\mathbf{x}_i) \ d\boldsymbol\xi.
\end{align}
What is retained in the present approximation is the first term, the error thus reducing to the last term of the expansion (equate the Taylor expansion Eq.\eqref{eq:Taylor} with Eq.\eqref{eq:mass_approx_g}). The mass approximation is denoted by $P_i^{(G)}$ and it can be defined as:
\begin{equation}
P_i^{(G)} = P_i^{\mathrm{(KL)}} + \frac{p(\mathbf{x}_i)}{2 g(\mathbf{x}_i)} \int_{\mathscr{B}(\varepsilon,\mathbf{x}_{i})} (\boldsymbol{\xi}-\mathbf{x}_i)^T \left[\nabla \nabla g |_{\mathbf{x}_i} \right](\boldsymbol{\xi}-\mathbf{x}_i) \ d\boldsymbol\xi.
\end{equation}
Roughly speaking, the mass is the sum of the mass obtained by the KL hypothesis plus an additional term that results from the approximation of Hessian of the target distribution by means of the Hessian of the empirically estimated gaussian.

The error is denoted by $ e_{P_i}^{(G)}:=|P_i - P_i^{(G)}|$:
\begin{align}
e_{P_i}^{(G)} = \frac{1}{2} \int_{\mathscr{B}(\varepsilon,\mathbf{x}_{i})}(\boldsymbol{\xi}-\mathbf{x}_i)^T \left[\nabla \nabla R |_{\mathbf{x}_i} \right] (\boldsymbol{\xi}-\mathbf{x}_i) \ d\boldsymbol\xi.
\end{align}

If the distribution is Gaussian and it is perfectly estimated through the samples, this term vanishes. Remark that, the behaviour of the error as function of the dimension is exactly the same as for the KL estimator, but if the Hessian of the Gaussian estimates the Hessian of the target distribution, the upper bound on the error will be smaller.

\subsubsection{Error in the approximation of the entropy}
The error on the entropy estimate is computed by following exactly the same strategy as for the KL estimator. The upper and lower bounds have the same expression, except that the eigenvalues appearing (namely $\zeta_i^{min,max}$) in the expressions are those of the Hessian of the residual $R$.

The lower bound reads:
\begin{align}
| H - H^{(G)} |  \geq \left | e_S + \frac{d\ 2^{d-1}}{3N}\sum_i^N \frac{\zeta_i^{min}\varepsilon_i^{d+2}}{P_i^{\mathrm{(KL)}} +  \frac{| \zeta^{max}_i| d 2^{d-1}}{3} \varepsilon_i^{d+2} }  \right|.
\label{eq:G_lower_bound}
\end{align}

And the upper bound is:
\begin{align}
| H - H^{(G)} |  \leq e_S + \frac{d\ 2^{d-1}}{3N}\sum_i^N \frac{|\zeta_i^{max}|}{P_i^{\mathrm{(KL)}}} \varepsilon_i^{d+2}.
\label{eq:G_upper_bound}
\end{align}

\bibliography{references}

\begin{thebibliography}{15}%
\makeatletter
\providecommand \@ifxundefined [1]{%
 \@ifx{#1\undefined}
}%
\providecommand \@ifnum [1]{%
 \ifnum #1\expandafter \@firstoftwo
 \else \expandafter \@secondoftwo
 \fi
}%
\providecommand \@ifx [1]{%
 \ifx #1\expandafter \@firstoftwo
 \else \expandafter \@secondoftwo
 \fi
}%
\providecommand \natexlab [1]{#1}%
\providecommand \enquote  [1]{``#1''}%
\providecommand \bibnamefont  [1]{#1}%
\providecommand \bibfnamefont [1]{#1}%
\providecommand \citenamefont [1]{#1}%
\providecommand \href@noop [0]{\@secondoftwo}%
\providecommand \href [0]{\begingroup \@sanitize@url \@href}%
\providecommand \@href[1]{\@@startlink{#1}\@@href}%
\providecommand \@@href[1]{\endgroup#1\@@endlink}%
\providecommand \@sanitize@url [0]{\catcode `\\12\catcode `\$12\catcode
  `\&12\catcode `\#12\catcode `\^12\catcode `\_12\catcode `\%12\relax}%
\providecommand \@@startlink[1]{}%
\providecommand \@@endlink[0]{}%
\providecommand \url  [0]{\begingroup\@sanitize@url \@url }%
\providecommand \@url [1]{\endgroup\@href {#1}{\urlprefix }}%
\providecommand \urlprefix  [0]{URL }%
\providecommand \Eprint [0]{\href }%
\providecommand \doibase [0]{http://dx.doi.org/}%
\providecommand \selectlanguage [0]{\@gobble}%
\providecommand \bibinfo  [0]{\@secondoftwo}%
\providecommand \bibfield  [0]{\@secondoftwo}%
\providecommand \translation [1]{[#1]}%
\providecommand \BibitemOpen [0]{}%
\providecommand \bibitemStop [0]{}%
\providecommand \bibitemNoStop [0]{.\EOS\space}%
\providecommand \EOS [0]{\spacefactor3000\relax}%
\providecommand \BibitemShut  [1]{\csname bibitem#1\endcsname}%
\let\auto@bib@innerbib\@empty
\bibitem [{\citenamefont {Cover}\ and\ \citenamefont
  {Thomas}(2012)}]{cover2012elements}%
  \BibitemOpen
  \bibfield  {author} {\bibinfo {author} {\bibfnamefont {T.~M.}\ \bibnamefont
  {Cover}}\ and\ \bibinfo {author} {\bibfnamefont {J.~A.}\ \bibnamefont
  {Thomas}},\ }\href@noop {} {\emph {\bibinfo {title} {Elements of information
  theory}}}\ (\bibinfo  {publisher} {John Wiley \& Sons},\ \bibinfo {year}
  {2012})\BibitemShut {NoStop}%
\bibitem [{\citenamefont {Beirlant}\ \emph {et~al.}(1997)\citenamefont
  {Beirlant}, \citenamefont {Dudewicz}, \citenamefont {Gy{\"o}rfi},\ and\
  \citenamefont {Van~der Meulen}}]{beirlant1997nonparametric}%
  \BibitemOpen
  \bibfield  {author} {\bibinfo {author} {\bibfnamefont {J.}~\bibnamefont
  {Beirlant}}, \bibinfo {author} {\bibfnamefont {E.~J.}\ \bibnamefont
  {Dudewicz}}, \bibinfo {author} {\bibfnamefont {L.}~\bibnamefont
  {Gy{\"o}rfi}}, \ and\ \bibinfo {author} {\bibfnamefont {E.~C.}\ \bibnamefont
  {Van~der Meulen}},\ }\href@noop {} {\bibfield  {journal} {\bibinfo  {journal}
  {Int. J. Math. Stat. Sci.}\ }\textbf {\bibinfo {volume} {6}},\ \bibinfo
  {pages} {17} (\bibinfo {year} {1997})}\BibitemShut {NoStop}%
\bibitem [{\citenamefont {Silverman}(1986)}]{silverman1986density}%
  \BibitemOpen
  \bibfield  {author} {\bibinfo {author} {\bibfnamefont {B.~W.}\ \bibnamefont
  {Silverman}},\ }\href@noop {} {\emph {\bibinfo {title} {Density estimation
  for statistics and data analysis}}},\ Vol.~\bibinfo {volume} {26}\ (\bibinfo
  {publisher} {CRC press},\ \bibinfo {year} {1986})\BibitemShut {NoStop}%
\bibitem [{\citenamefont {Devroye}\ and\ \citenamefont
  {Gy{\"o}rfi}(1985)}]{devroye1985nonparametric}%
  \BibitemOpen
  \bibfield  {author} {\bibinfo {author} {\bibfnamefont {L.}~\bibnamefont
  {Devroye}}\ and\ \bibinfo {author} {\bibfnamefont {L.}~\bibnamefont
  {Gy{\"o}rfi}},\ }\href {https://books.google.fr/books?id=ew7vAAAAMAAJ} {\emph
  {\bibinfo {title} {Nonparametric density estimation: the L1 view}}},\ Wiley
  series in probability and mathematical statistics\ (\bibinfo  {publisher}
  {Wiley},\ \bibinfo {year} {1985})\BibitemShut {NoStop}%
\bibitem [{\citenamefont {Scott}(2015)}]{scott2015multivariate}%
  \BibitemOpen
  \bibfield  {author} {\bibinfo {author} {\bibfnamefont {D.~W.}\ \bibnamefont
  {Scott}},\ }\href@noop {} {\emph {\bibinfo {title} {Multivariate density
  estimation: theory, practice, and visualization}}}\ (\bibinfo  {publisher}
  {John Wiley \& Sons},\ \bibinfo {year} {2015})\BibitemShut {NoStop}%
\bibitem [{\citenamefont {Hall}(1984)}]{hall1984limit}%
  \BibitemOpen
  \bibfield  {author} {\bibinfo {author} {\bibfnamefont {P.}~\bibnamefont
  {Hall}},\ }in\ \href@noop {} {\emph {\bibinfo {booktitle} {Mathematical
  Proceedings of the Cambridge Philosophical Society}}},\ Vol.~\bibinfo
  {volume} {96}\ (\bibinfo {organization} {Cambridge Univ Press},\ \bibinfo
  {year} {1984})\ pp.\ \bibinfo {pages} {517--532}\BibitemShut {NoStop}%
\bibitem [{\citenamefont {Dudewicz}\ and\ \citenamefont {{Van der
  Meulen}}(1987)}]{dudewicz1987empiric}%
  \BibitemOpen
  \bibfield  {author} {\bibinfo {author} {\bibfnamefont {E.}~\bibnamefont
  {Dudewicz}}\ and\ \bibinfo {author} {\bibfnamefont {E.}~\bibnamefont {{Van
  der Meulen}}},\ }in\ \href@noop {} {\emph {\bibinfo {booktitle} {New
  Perspectives in Theoretical and Applied Statistics}}},\ \bibinfo {editor}
  {edited by\ \bibinfo {editor} {\bibfnamefont {W.~W.~E.}\ \bibnamefont
  {M.L.~Puri}, \bibfnamefont {J.~Vilaplana}}}\ (\bibinfo  {publisher} {Wiley,
  New York},\ \bibinfo {year} {1987})\BibitemShut {NoStop}%
\bibitem [{\citenamefont {Kozachenko}\ and\ \citenamefont
  {Leonenko}(1987)}]{Kozachenko1987sample}%
  \BibitemOpen
  \bibfield  {author} {\bibinfo {author} {\bibfnamefont {L.~F.}\ \bibnamefont
  {Kozachenko}}\ and\ \bibinfo {author} {\bibfnamefont {N.~N.}\ \bibnamefont
  {Leonenko}},\ }\href@noop {} {\bibfield  {journal} {\bibinfo  {journal}
  {Probl. Inf. Transm.}\ }\textbf {\bibinfo {volume} {23}},\ \bibinfo {pages}
  {9} (\bibinfo {year} {1987})}\BibitemShut {NoStop}%
\bibitem [{\citenamefont {Tsybakov}\ and\ \citenamefont {Van~der
  Meulen}(1996)}]{tsybakov1996root}%
  \BibitemOpen
  \bibfield  {author} {\bibinfo {author} {\bibfnamefont {A.~B.}\ \bibnamefont
  {Tsybakov}}\ and\ \bibinfo {author} {\bibfnamefont {E.}~\bibnamefont {Van~der
  Meulen}},\ }\href@noop {} {\bibfield  {journal} {\bibinfo  {journal}
  {Scandinavian Journal of Statistics}\ ,\ \bibinfo {pages} {75}} (\bibinfo
  {year} {1996})}\BibitemShut {NoStop}%
\bibitem [{\citenamefont {Singh}\ \emph {et~al.}(2003)\citenamefont {Singh},
  \citenamefont {Misra}, \citenamefont {Hnizdo}, \citenamefont {Fedorowicz},\
  and\ \citenamefont {Demchuk}}]{singh2003nearest}%
  \BibitemOpen
  \bibfield  {author} {\bibinfo {author} {\bibfnamefont {H.}~\bibnamefont
  {Singh}}, \bibinfo {author} {\bibfnamefont {N.}~\bibnamefont {Misra}},
  \bibinfo {author} {\bibfnamefont {V.}~\bibnamefont {Hnizdo}}, \bibinfo
  {author} {\bibfnamefont {A.}~\bibnamefont {Fedorowicz}}, \ and\ \bibinfo
  {author} {\bibfnamefont {E.}~\bibnamefont {Demchuk}},\ }\href@noop {}
  {\bibfield  {journal} {\bibinfo  {journal} {Amer. J. Math. Management Sci.}\
  }\textbf {\bibinfo {volume} {23}},\ \bibinfo {pages} {301} (\bibinfo {year}
  {2003})}\BibitemShut {NoStop}%
\bibitem [{\citenamefont {Kraskov}\ \emph {et~al.}(2004)\citenamefont
  {Kraskov}, \citenamefont {St\"ogbauer},\ and\ \citenamefont
  {Grassberger}}]{kraskov2004estimating}%
  \BibitemOpen
  \bibfield  {author} {\bibinfo {author} {\bibfnamefont {A.}~\bibnamefont
  {Kraskov}}, \bibinfo {author} {\bibfnamefont {H.}~\bibnamefont
  {St\"ogbauer}}, \ and\ \bibinfo {author} {\bibfnamefont {P.}~\bibnamefont
  {Grassberger}},\ }\href {\doibase 10.1103/PhysRevE.69.066138} {\bibfield
  {journal} {\bibinfo  {journal} {Phys. Rev. E}\ }\textbf {\bibinfo {volume}
  {69}},\ \bibinfo {pages} {066138} (\bibinfo {year} {2004})}\BibitemShut
  {NoStop}%
\bibitem [{\citenamefont {Gray}\ and\ \citenamefont
  {Moore}(2003)}]{gray2003nonparametric}%
  \BibitemOpen
  \bibfield  {author} {\bibinfo {author} {\bibfnamefont {A.~G.}\ \bibnamefont
  {Gray}}\ and\ \bibinfo {author} {\bibfnamefont {A.~W.}\ \bibnamefont
  {Moore}},\ }in\ \href@noop {} {\emph {\bibinfo {booktitle} {SDM}}}\ (\bibinfo
  {organization} {SIAM},\ \bibinfo {year} {2003})\ pp.\ \bibinfo {pages}
  {203--211}\BibitemShut {NoStop}%
\bibitem [{\citenamefont {Gao}\ \emph {et~al.}(2015)\citenamefont {Gao},
  \citenamefont {Steeg},\ and\ \citenamefont {Galstyan}}]{gao2014Efficient}%
  \BibitemOpen
  \bibfield  {author} {\bibinfo {author} {\bibfnamefont {S.}~\bibnamefont
  {Gao}}, \bibinfo {author} {\bibfnamefont {G.~V.}\ \bibnamefont {Steeg}}, \
  and\ \bibinfo {author} {\bibfnamefont {A.}~\bibnamefont {Galstyan}},\ }\href
  {http://arxiv.org/abs/1411.2003} {\bibfield  {journal} {\bibinfo  {journal}
  {{Arxiv (http://arxiv.org/abs/1411.2003)}}\ } (\bibinfo {year}
  {2015})}\BibitemShut {NoStop}%
\bibitem [{\citenamefont {Genz}(1992)}]{genz1992numerical}%
  \BibitemOpen
  \bibfield  {author} {\bibinfo {author} {\bibfnamefont {A.}~\bibnamefont
  {Genz}},\ }\href@noop {} {\bibfield  {journal} {\bibinfo  {journal} {J. Comp.
  Graph. Stat.}\ }\textbf {\bibinfo {volume} {1}},\ \bibinfo {pages} {141}
  (\bibinfo {year} {1992})}\BibitemShut {NoStop}%
\bibitem [{\citenamefont {Cunningham}\ \emph {et~al.}(2012)\citenamefont
  {Cunningham}, \citenamefont {Hennig},\ and\ \citenamefont
  {Lacoste-Julien}}]{CunninghamHL2012}%
  \BibitemOpen
  \bibfield  {author} {\bibinfo {author} {\bibfnamefont {J.}~\bibnamefont
  {Cunningham}}, \bibinfo {author} {\bibfnamefont {P.}~\bibnamefont {Hennig}},
  \ and\ \bibinfo {author} {\bibfnamefont {S.}~\bibnamefont {Lacoste-Julien}},\
  }\href@noop {} {\bibfield  {journal} {\bibinfo  {journal} {{Arxiv
  (http://arxiv.org/abs/1111.6832)}}\ } (\bibinfo {year} {2012})}\BibitemShut
  {NoStop}%
\end{thebibliography}%

\end{document}